\documentclass{ws-ijmpb2}

\begin{document}

\markboth{C. Ligi et al.}
{Measurement of the thermal expansion}

%
\catchline{}{}{}{}{}
%

\title{MEASUREMENT OF THE THERMAL EXPANSION COEFFICIENT OF AN Al-Mg ALLOY AT ULTRA-LOW TEMPERATURES}

\author{M.~BASSAN$^{b,c}$, B.~BUONOMO$^a$, G.~CAVALLARI$^e$, E.~COCCIA$^{b,c}$, S.~D'ANTONIO$^b$, V.~FAFONE$^{b,c}$, L.G.~FOGGETTA$^a$, C.~LIGI$^a$\footnote{Corresponding author} , A.~MARINI$^a$, G.~MAZZITELLI$^a$, G.~MODESTINO$^a$, G.~PIZZELLA$^{c,a}$, L.~QUINTIERI$^a$\footnote{Present address: ENEA C. R. Casaccia, Via Anguillarese 301, I-00123 Roma, Italy} , F.~RONGA$^a$, P.~VALENTE$^d$}

\address{$^a$Istituto Nazionale di Fisica Nucleare - Laboratori Nazionali di Frascati,\\ Via~E.~Fermi,~40~- 00044 Frascati, Italy\\
$^b$Istituto Nazionale di Fisica Nucleare - Sezione Roma2,\\ Via~della~Ricerca~Scientifica~- 00133 Rome, Italy\\
$^c$Dipartimento di Fisica, Universit\`a di Tor Vergata,\\ Via~della~Ricerca~Scientifica~- 00133 Rome, Italy\\
$^d$Istituto Nazionale di Fisica Nucleare - Sezione Roma1,\\ Piazzale Aldo Moro 2~- 00185 Rome, Italy\\
$^e$CERN, CH1211, Gen\`eve, Switzerland\\
carlo.ligi@lnf.infn.it}

\maketitle


\begin{abstract}
We describe a result coming from an experiment based on an Al-Mg alloy ($\sim 5\% $ Mg) suspended bar hit by an electron beam and operated above and below the termperature of transition from superconducting to normal state of the material. The amplitude of the bar first longitudinal mode of oscillation, excited by the beam interacting with the bulk, and the energy deposited by the beam in the bar are the quantities measured by the experiment. These quantities, inserted in the equations describing the mechanism of the mode excitation and complemented by an independent measurement of the specific heat, allow us to determine the linear expansion coefficient $\alpha$ of the material. We obtain 
$\alpha=[(10.9\pm0.4)T+(1.3\pm0.1)T^3]\times 10^{-10}\ \rm{K^{-1}}$ for the normal state of conduction  in the temperature interval $0.9< T< 2$~K and $\alpha=[(-2.45\pm0.60)+(-10.68\pm1.24)T +(0.13\pm0.01)T^3]\times 10^{-9}\ \rm{K^{-1}}$ for the superconducting state in the interval $0.3 <T<0.8$~K.
\end{abstract}

\keywords{Thermal expansion; Low temperature; Aluminum alloy.}


\section{Introduction}
\label{intro}

Wrought aluminum-magnesium alloys (International Alloy Designation System: 5XXX, Mg content: 5.6\% maximum)  are commonly used in applications where good workability, very good resistance to corrosion, high fatigue strength are desidered. Example of applications are: oil, fuel lines and tanks, pressure cryogenic vessels, marine structures and fittings, automotive and architectural components. 

The alloy Al~5056 (5.2\% Mg,  0.1\% Mn, 0.1\% Cr) is the material of the test mass of  gravitational wave (GW) resonant antennas AURIGA\cite{auriga}, EXPLORER\cite{explorer}, NAUTILUS\cite{rog} in Italy and ALLEGRO\cite{allegro} in the United States, operated at liquid Helium temperature and below. During the last decade, all these detectors took part, as a network, in coordinated searches for impulsive GW excitations (no GW events were detected).\cite{igec1,igec2}

The test mass of the GW resonant antennas is a suspended cylinder ($\sim$~2300 kg mass), acting as a mechanical resonator whose resonances are excited by the incoming GW. The elastic vibrations are converted to electrical signals by a transducer system. The minimum detectable energy of a GW resonant antenna  inversely depends on $Q$, the acoustic quality factor of the material  also known as  the inverse of the internal friction. Al~5056 was chosen as the material for the GW antennas due to the high values of $Q$,\cite{Q} exhibited at very low temperatures. 

B.L.~Baron and R.~Hofstadter first measured\cite{beron} mechanical oscillations in piezoelectric disks penetrated by high energy electron beams and they outlined the possibility that cosmic ray (CR) events could excite mechanical vibrations in a metallic cylinder at its resonant frequency and that these events could mimic GW events detected by resonant antennas. NAUTILUS has been equipped with a CR detector\cite{streamer} to study the interactions caused by CRs and to provide a veto against the CR induced events in the antenna.

\noindent Correlating NAUTILUS data with CR observations, the following results were obtained: 1)  the rate of high energy signals due to CR showers was larger than expectations\cite{naut1,naut2} when the antenna was operated at a temperature $T=0.14\ \rm{K}$, {\it i.e.}~well below the transition temperature from normal-conducting $(n)$ to superconducting $(s)$ states of Al~5056;\cite{coccia} 2) this feature was not observed when the antenna was operated at $T=1.5\ \rm{K}$, {\it i.e.}~above the transition temperature.\cite{naut3} \\
Since these findings had not a straightforward interpretation, we performed an experiment (RAP) aimed at measuring the longitudinal oscillations of a small suspended Al~5056 cylinder hit by an electron beam of known energy.
The RAP experiment, performed at the Beam Test Facility (BTF)\cite{mazbtf} of the DAFNE $\Phi$-factory complex in the INFN Frascati Laboratory, has shown  that 1) the amplitude of the longitudinal oscillations of the cylinder hit by ionizing particles depends on the state of conduction of the material\cite{al2,nb} and that 2)  the observed  amplitudes are consistent with the amplitudes measured by NAUTILUS in CR events both in $(n)$ and $(s)$ states.\cite{al3} Moreover, the amplitude can be evaluated in the framework of the Thermo-Acoustic model: in the $(n)$ state it depends on the ratio of the thermal expansion coefficient to the specific heat, this ratio being part of the definition of the Gr\"{u}neisen parameter. In the $(s)$ state the amplitude also depends on the fractional volume change between the $(s)$ and $(n)$ states. 
In principle,  the relevant thermophysical properties of the Al~5056 at low temperature are not known with sufficient precision to allow the comparison between experimental and theoretical values of the amplitude predicted by the model. However, our confidence in the model is also based on the small discrepancy, of the order of 15\%, between the amplitudes measured in the alloy and those computed by using the data for pure aluminum in the region 4.5$\ \leq T\leq$ 264~K.\cite{al1}

Direct measurements of the Gr\"{u}neisen parameter of elastic materials have been reported:\cite{barron} the authors used pulsed electron beams hitting the front surface of a thin slab, and measured the induced motion on the rear face surface. The equation of motion describing thermoelastic effects depends on the Gr\"{u}neisen parameter and on the energy deposited by the beam in the slab. The measurement of the displacement, the knowledge of the deposited energy and the use of the equation of motion allowed them to directly determine the Gr\"{u}neisen parameter.

\par\noindent In a similar way this paper presents experimental results on  the  linear expansion coefficient of the aluminum alloy  above and below the temperature of transition between the $(s)$ and $(n)$ state. In order to determine the coefficient we use: a) the oscillation amplitude measurements, b) the equations describing the amplitude in the framework of the Thermo-Acoustic model, c) the measurements of the deposited energy by the electron beam, obtained by the product of the measured beam multiplicity and the calculated energy loss in the material by each particle, and d) an independent measurement of the specific heat. 
\section{The Thermo-Acoustic model}
\label{TAM}

\subsection{$(n)$ state}
\label{nstate}

We consider a cylinder (radius $R$, length $L${, mass $M$}) that is suspended in correspondence of the middle section, with its axis horizontal. A ionizing particle, after hitting the lateral surface of the cylinder and interacting with the material, generates a pressure pulse in the bulk. This sonic pulse is the result of the local thermal expansion of the material caused by the warming up due to the energy lost by the particle in the material. The sonic pulse determines the excitation of the vibrational oscillation modes of the suspended bar. By introducing a vector field $\mathbf{u}({\mathbf{x}},t)$ to describe the local displacements from equilibrium, we express the amplitude of the $k$-th longitudinal mode of oscillation  as proportional to the quantity:\cite{allega,deru,liu}
\begin{eqnarray}
g_{k,therm}& = & {\frac{\Delta P_{therm}}{\rho}} {\cal A'} {\cal I}{_k} \nonumber \\
                          & =  & {\frac{\gamma}{\rho}}\left| {\frac{dW}{dl}}\right|  {\cal I}{_k} \ ,
\label{gtherm}
\end{eqnarray}
\noindent where  $\Delta P_{therm}$ is the pressure pulse due to the sonic source described above, $\rho$ is the mass density, $dW/dl$ is the energy loss per unit length of the interacting particle,
$\cal A'$ is the cross section of the tubular zone centered on the particle path in which the effects are generated and ${\cal I}{_k}=\int{dl (\nabla\cdot\mathbf{u}{_k}({\mathbf{x}}))}$ is a line integral over the particle path involving the normal mode of oscillation $\mathbf{u}{_k}({\mathbf{x}})$. In the previous relation $\gamma$ is the adimensional Gr\"{u}neisen parameter:
\begin{equation}
\gamma=\frac{\beta K_{T}}{\rho c_{V}}\ ,
\label{gru}
\end{equation}
\noindent with $\beta$  the volume thermal expansion coefficient  ($\beta=3\alpha$ for cubic elements), $c_V$ the isochoric specific heat and $K_{T}$ the isothermal bulk module. 

A thin bar $(R/L\ll 1)$ hit by a ionizing particle at the center of the lateral surface is a particular case of the more general one that leads to relation~(\ref{gtherm}).
In this case, the fundamental longitudinal mode of oscillation is excited to a maximum amplitude given by:\cite{strini}

\begin{equation}
X=\frac{2 \alpha L W}{\pi c_V M}\ ,
\label{b0}
\end{equation}
and then exponentially decays due to internal friction. Here $W$ is the total energy released in the bar by the particle beam. The experimental conditions of RAP are close to this particular case and the amplitude maximum  of the fundamental longitudinal mode of oscillation is the observable measured by the experiment. However, in order to model the amplitude in the most realistic way, we have performed  a Monte Carlo (MC) simulation,\cite{al1} which takes into account the corrections $O[(R/L)^2]$ for the modes of oscillation of a cylinder, the transverse dimension of the electron beam at the impact point  and the trajectories of the secondary particles generated by the electron interactions in the bar. All these effects are summarized in a  corrective parameter $\epsilon$ and the amplitude maximum of the fundamental longitudinal mode of oscillation is modeled by:
 
\begin{equation}
X_{(n)}=X(1+\epsilon)   
\label{xth}
\end{equation}
    
\par\noindent The value of  $\epsilon$ for the aluminum alloy bar used in the experiment is estimated by MC to be -0.04.

\subsection{$(s)$ state}
\label{sstate}

The energy released by the ionizing particles to the material determines the suppression of the superconductivity in a region (Hot Spot) centered around the particle path.
The maximum value of the Hot Spot radius, $r_{HS}$, is obtained by equating the energy lost per unit length by the particle, $dW/dl$, to the enthalpy variation in the volume undergoing the $(s)\rightarrow (n)$ transition at temperature $T$:\cite{gray,strehl,lisitskiy}
\begin{equation}
r_{HS}  = \sqrt{\frac{|dW/dl|}{\pi\ \Delta h}} =  \sqrt{\frac{\cal A
''}{\pi}} \ ,
\label{rHS}
\end{equation}
\noindent  where  $\Delta h$ is the enthalpy change per unit volume and $\cal A''$ is the cross section of the zone switched to the $(n)$ state. 

\noindent The amplitude of the longitudinal oscillations is the sum of two terms, one related to thermodynamic effects in the $(s)\rightarrow (n)$ transition and the other to the thermal effects in the region switched to the $(n)$ state, which have been already described in the Section~\ref{nstate}. The  contribution to the amplitude of the cylinder $k-th$ oscillation mode  due to the  $(s)\rightarrow (n)$ transition is proportional to:\cite{allega,deru}
\begin{eqnarray*}
g_{k,trans}& = & {\frac{\Delta P_{trans}}{\rho}} {\mathcal A''} {\mathcal I}{_k}  \\
& =  & {\frac{1}{\rho}}\left[{ K_{T}\frac{\Delta V}{V}+\gamma T \frac{\Delta \mathcal{S}}{V}}\right]  {\mathcal A''}{\mathcal I}{_k} \ ,
\end{eqnarray*}

\noindent where $V=V_{(s)}$ and $\Delta V$, $\Delta \mathcal{S}$ are the differences of the volume and the entropy in the two states of conduction. According to the thermodynamics of volume and pressure effects for Type-II superconductors, these differences  can be written as\footnote[1]{We show the relations in $cgs$ unit system, as done by the original authors. In the computations we will convert the magnetic field energy density from $cgs$ to $SI$. }:\cite{hake}
\begin{equation}
\Delta V=V_{(n)}- V_{(s)}=\frac{VH_c}{4\pi}\frac{\partial H_c}{\partial P} +\left\{ \frac{H_c^2}{8\pi}\frac{\partial V}{\partial P}\right\}
\label{Dvolume}
\end{equation}

\noindent and

\begin{equation}
\Delta {\mathcal S}={\mathcal S}_{(n)}- {\mathcal S}_{(s)}=-\frac{VH_c}{4\pi}\frac{\partial H_c}{\partial T} - \left\{ \frac{H_c^2}{8\pi}\frac{\partial V}{\partial T}\right\}
\label{Dentropy}
\end{equation}

\noindent Here $H_c$ is the superconducting critical magnetic field that is supposed to have the parabolic behavior $H_c(t)=H_c(0)(1-t^2)$, where $t=T/T_c$ and $T_c$ is the transition temperature. Moreover, the terms in $\{\ \}$ brackets are expected to be  smaller than the preceding terms\cite{shoenberg} and are usually ignored for practical purposes.

\noindent Finally, the knowledge of the specific heat of the material for the $(s)$ state, $c^{(s)}$, allows us to compute the variation of the entalpy per unit volume in the form:\cite{whited}

\begin{equation}
\Delta h=C{_{int}}(T)+T\frac{\Delta{\mathcal S}}{V}
\label{Dentalpy}
\end{equation}

\noindent with:
\begin{displaymath}
C{_{int}}(T)=\int_T^{T_c}c^{(s)}(T') dT'
\end{displaymath}

The amplitude maximum  of the fundamental longitudinal mode of oscillation in the $(s)$ state, as already indicated, is the sum of two terms, one related to the transition and the other to the $(n)$ contribution from the switched region: $X_{(s)}=X_{trans}+X_{(n)}$. By observing that $X_{trans}/X_{(n)}=g{_{k=0,trans}}/g{_{k=0,therm}}$ we can describe the amplitude maximum according to the following relation:

\begin{eqnarray}
\lefteqn{X_{(s)}= X_{(n)} \left\{ 1 + \frac{X_{trans}}{ X_{(n)}}\right\}} \nonumber \\
& =  & X_{(n)} \left\{1 + \left[ \Pi \frac{\Delta V}{V}+T \frac{\Delta \mathcal{S}}{V} \right] \left[{ \Delta h}\right]^{-1} \right\}\ ,
\label{exp}
\end{eqnarray}

\noindent where the relation (\ref{xth}) and the definition (\ref{gru}) of  $\gamma$ are used to define:
\begin{displaymath}
\Pi = \frac{2\rho L (1+\epsilon)}{3\pi M (X_{(n)}/W)}
\end{displaymath}

\noindent We note that  the term contained  in $\{\ \}$ brackets in the relation~(\ref{exp}) is independent from $W$, the energy released by the particle to the bar, and that $X_{(s)}$ linearly depends on $W$ through $X_{(n)}$.

\section{Experimental setup and procedures}
\label{exset}
The experiment mechanical layout, its cryogenic setup and operations, the electron beam characteristics, the instrumentation and the procedures for calibrations, data taking and analysis have been fully described in Refs.~17--18. Briefly, the dimensions and the mass of the Al~5056 cylindrical bar are $R=0.091\ \rm{m}$, $L=0.500\ \rm{m}$, $M=34.1\ \rm{kg}$, respectively.  The bar hangs from the cryostat top by means of a multi-stage suspension system providing a 150 dB attenuation of the external mechanical noise  in the 1700-6500~Hz frequency window. The frequency of the fundamental longitudinal mode of oscillation of the bar is $f_0=5413.6$~Hz below $T = 4$~K.  The cryostat is equipped with a dilution refrigerator. The temperatures are measured inside the cryostat by 11 thermometers controlled by two multi-channel resistance bridges and, among them, a calibrated $\rm{RuO_2}$ resistor measures the temperature of one of the bar end faces with an accuracy of  0.01~K below $T = 4$~K. Two piezoelectric ceramics (Pz), electrically connected in parallel, are inserted in a slot cut in the position opposite to the bar suspension point and are squeezed when the bar shrinks. In this Pz arrangement the strain measured at the bar center is proportional to the displacement of the bar end faces. The Pz output is first amplified and then sampled at 100~kHz by an ADC embedded in a VME system dedicated to the data acquisition.  A Pz calibration procedure, performed before each run of data taking, provides the factor converting the ADC samples into the displacements of the bar end faces. A software filtering algorithm, known as "digital lock-in", extracts the Fourier component at the frequency $f_0$ from the time sequence formed by the ADC samples before and after the beam hit, determining the amplitude of the induced fundamental oscillation. The sign of the amplitude is taken positive or negative according to the sign of the first sample raising above the threshold after the beam hit.

\noindent The BTF beam line  delivers to the bar single  pulses of $\sim10$~ns duration,  containing $N_e$ electrons of $510\pm 2$~MeV energy.  $N_e$ ranges from about $5\times 10^7$ to $10^9$ and is measured with an accuracy of $\sim 3 \%$  (for $N_e > 5\times 10^8$) by an integrating current transformer placed close to the beam line exit point. The MC simulation, discussed in Section~\ref{TAM}, estimates an average energy lost $\langle {\Delta  E}\rangle \pm \  \sigma_{\Delta  E}=195.2 \pm 70.6 \ \rm{MeV}$ for a 512~MeV electron interacting in the bar and, consequently, the total energy loss per beam pulse is given by $W=N{_e} \langle {\Delta  E}\rangle, \ \sigma_{W}=\sqrt{N{_e}}\  \sigma_{\Delta E}$. Two sources of error affect the vibration maximum amplitude: the first is an overall systematic error of the order of $\pm 6\%$, that accounts
for the slightly different set-up and calibration procedures implemented in the runs over two years and the second is related to the  noise in the measurement of the oscillation amplitude ($\pm 1.3\times 10^{-13}$~m).

\section{Linear expansion coefficient measurements}
\label{alpha}

The coefficient $\alpha$ is determined by inserting the measured values of  the amplitude maximum  of the fundamental longitudinal mode of oscillation $(X_{(n)},X_{(s)})$ and the corresponding measurements of the deposited energy $(W)$ into the relations modeling the amplitudes (equations (\ref{xth}),(\ref{exp})). The measurement\cite{barcv} at very low temperatures of the specific heat of an Al~5056 sample, of the same production batch of the RAP bar, is a fundamental ingredient for the coefficient determination both in $(n)$ and $(s)$ states. Moreover, the superconducting characterization of the material has shown that $T_c=0.845\pm0.002$~K with a total transition width of about 0.1~K.\cite{barcv} Thus we determine $\alpha$ for the two states of conduction $(\alpha_{(n)}, \alpha_{(s)})$ according to this temperature value.

\subsection{$(n)$ state}
\label{anstate}

\begin{figure}[h]
\begin{center}
\includegraphics[width=1.0\linewidth]{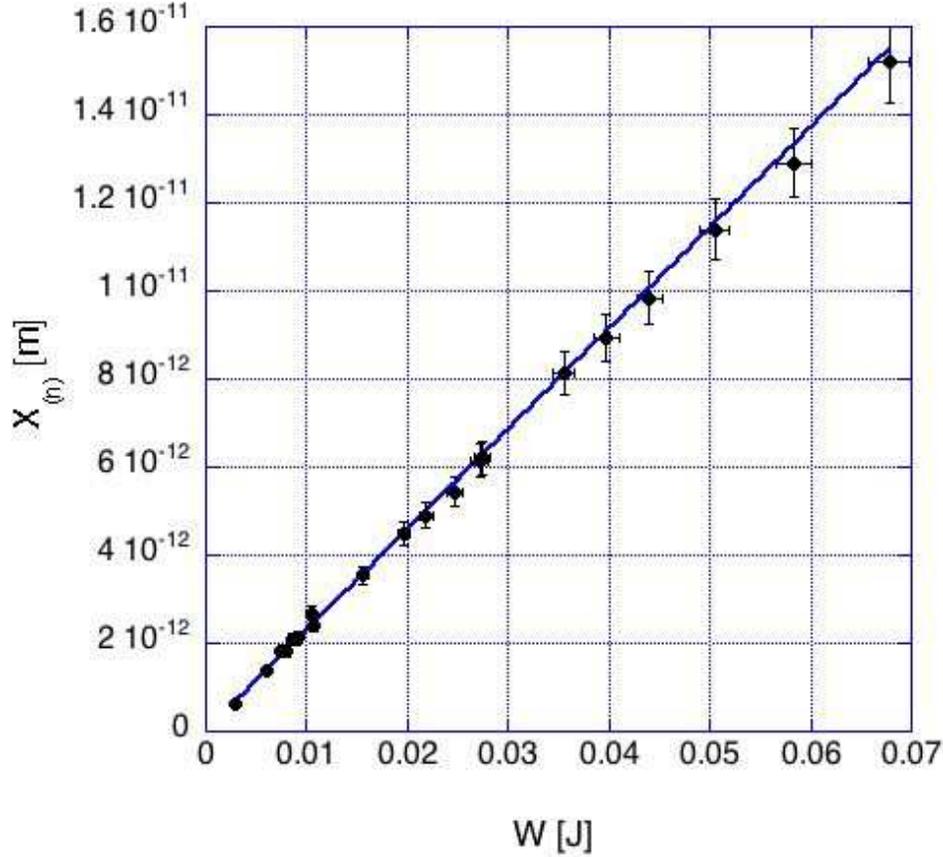}
\caption{\it Normal state. Values of the amplitude maximum  of the fundamental longitudinal mode of oscillation $(X_{(n)})$ vs.~the deposited energies (W) measured in the temperature interval $0.9\leq T\leq 2$ \rm{K}\it. Data have been fitted by a straight line, constrained to zero, with slope $(2.24 \pm 0.05)\times10^{-10}\ \rm{m\ J^{-1}}$.}
\label{XvWfig}
\end{center}
\end{figure}

Values of the amplitude maximum of the fundamental longitudinal mode of oscillation and the related values of the deposited energies by the beam have been measured and analyzed in the temperature interval $0.9\le T\le 2$~K.\cite{al3} The values of $X_{(n)}$  show a strict linear correlation with the values of $W$ (Fig.~\ref {XvWfig}) and they have been fitted, including the errors, by a line constrained to the origin $(X_{(n)}=p_0 W)$. The slope of the fit, which has a $\chi^2$ per degree of freedom equal to $0.45 / 15$, is $p_{0}=(2.24 \pm 0.05)\times10^{-10}\ \rm{m\ J^{-1}}$. From the relation~(\ref{xth}) it follows that:

\begin{equation}
\frac{\alpha}{c_V}=\frac{\pi M p_0}{2 L (1+\epsilon)}
\label{aoc}
\end{equation}

\begin{figure}[h]
\begin{center}
\includegraphics[width=1.0\linewidth]{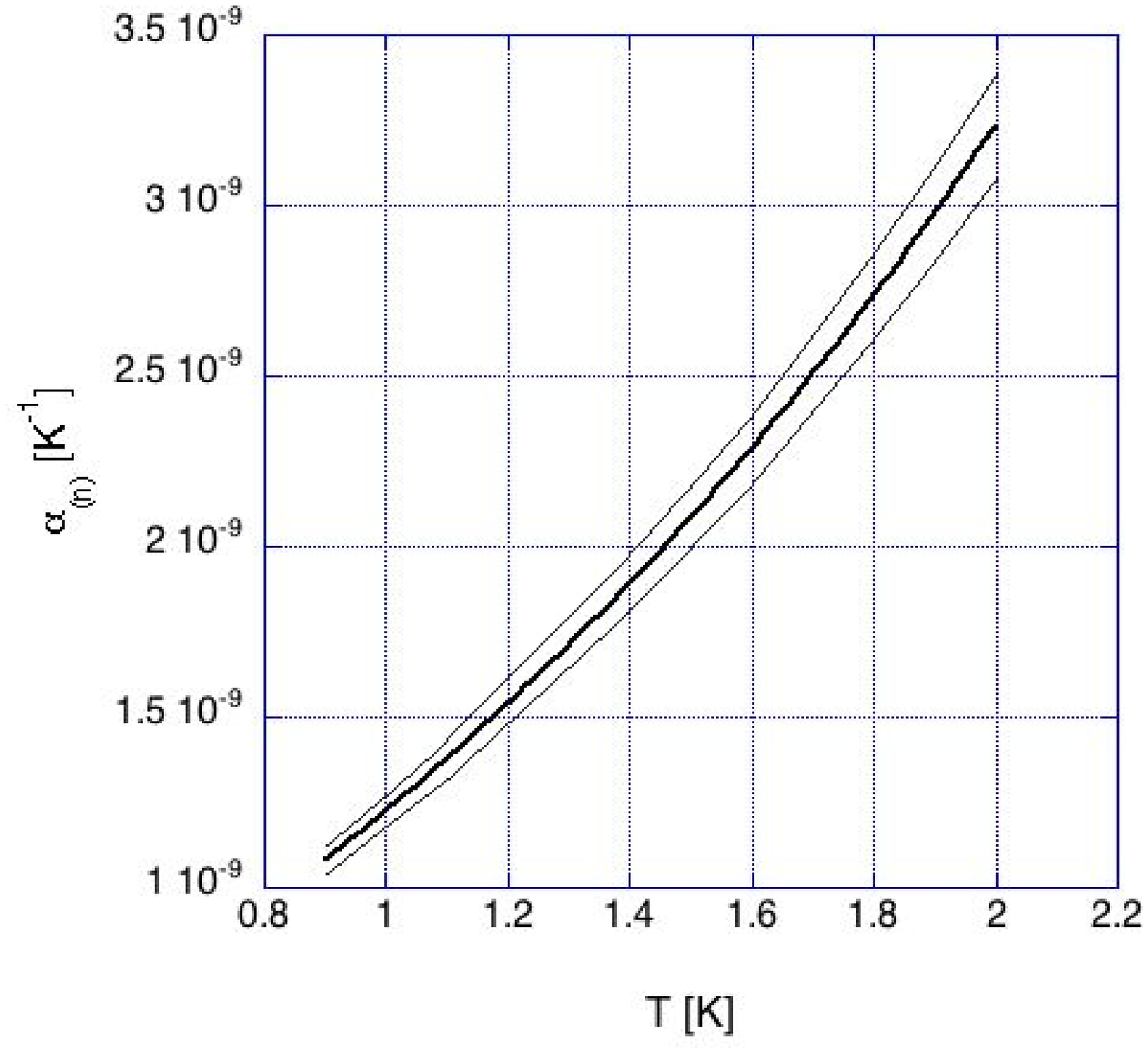}
\caption{\it Normal state. Measured values of the linear expansion coefficient ($\alpha_{(n)}$) vs.~temperature (T). The thinner lines enclose the region compatible with the uncertainties.}
\label{anfig}
\end{center}
\end{figure}

\noindent In the same temperature interval the parametrization  $c_V=CT+DT^3$, where  $C$ is the electronic specific heat coefficient per unit volume and $D$ is the lattice contribution, is determined by\cite{barcv}  $C=(4.382\pm 0.117)\times 10^{-2}\ \rm{J\ kg^{-1}\ K^{-2}}$ and $D=(5.20\pm 0.37)\times 10^{-3}\ \rm{J\ kg^{-1}\ K^{-4}}$.
Inserting this $c_V$ parametrization in relation (\ref{aoc}) directly gives:

\begin{eqnarray}
\lefteqn{\alpha_{(n)}=\alpha_{(n),e}+\alpha_{(n),\ell}} \nonumber \\
&=&[(10.9\pm0.4)T+(1.3\pm0.1)T^3]\times 10^{-10}\ \rm{K^{-1}}
\label{anmeas}
\end{eqnarray}

\noindent for the linear expansion coefficient (Fig.~\ref{anfig}) expressed in terms of the electronic and lattice components in the temperature interval $0.9\le T\le 2$~K.

\noindent We note that the value of $\alpha_{(n)}$ for pure aluminum is reported\cite{kroeger} to be $2.7\times 10^{-9}\ \rm{K^{-1}}$ at $T=2\ \rm{K}$. This value, which is the one associated to the lowest temperature in the previously cited work, can be compared with $\alpha_{(n)}=(3.2\pm 0.2)\times 10^{-9}\ \rm{K^{-1}}$ obtained in the present work for Al~5056 at the same temperature.

\subsection{$(s)$ state}
\label{asstate}

From eq.~(\ref{exp}) we can derive the volume variation among the two states of conduction:

\begin{equation}
\frac{\Delta V}{V}=\frac{1}{\Pi}\left[\frac{{\mathcal R}\ \Delta h}{p_0}-T\frac{\Delta{\mathcal S}}{V}\right]
\label{dvov}
\end{equation}

\noindent where ${\mathcal R}=X_{(s)}/W-p_0$, with $p_0$ introduced in Section~\ref{anstate}. The volume variation depends on measured quantities $(X_{(s)}, W, p_0, c_V,T,T_c)$ and calculated ones. The latter are related to the critical field $H_c(T=0)$ and to its parabolic dependance on the temperature, which also allows us to compute the derivative $\partial H_c/\partial T$. If the superconducting properties of Al~5056 can be described by the BCS theory, then $H_c(T=0) \approx 2.42 \sqrt{C} T_c \approx 70\ \rm{Oe}$, $C$ being the electronic specific heat coefficient per unit volume (Section~\ref{anstate}), here expressed in suitable $cgs$ units $(\rm{erg\ cm{^{-3}}\ K{^{-2}}})$.

\begin{figure}[htbp]
\begin{center}
\includegraphics[width=1.0\linewidth]{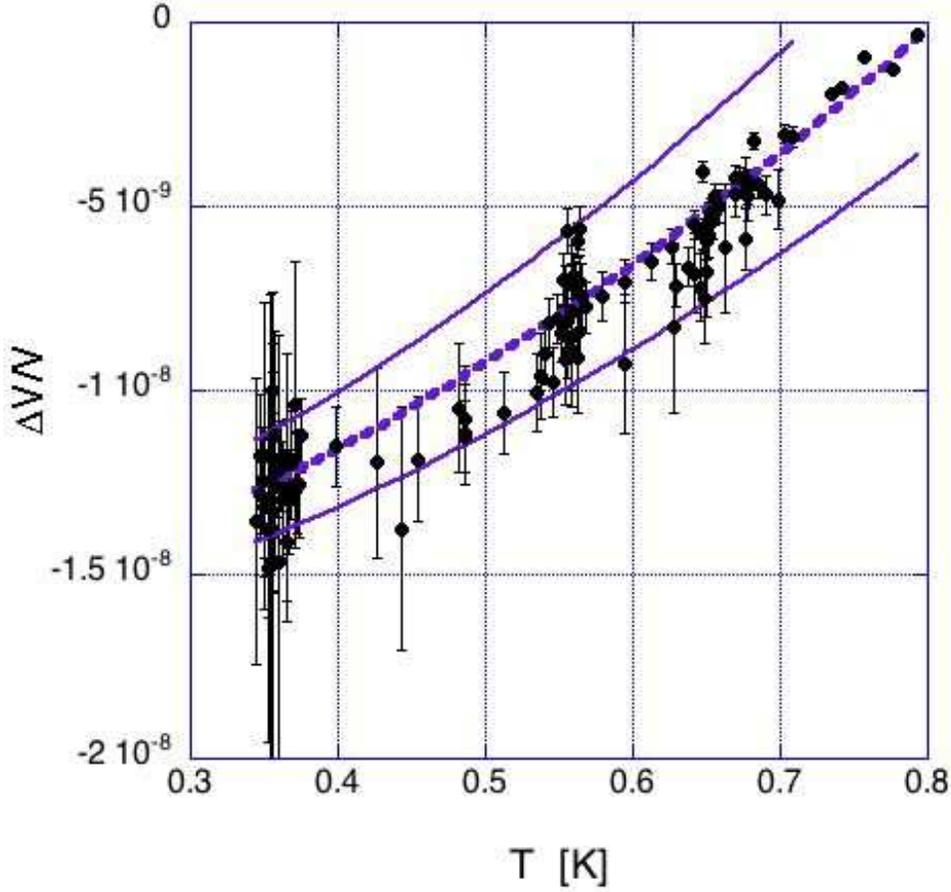}
\caption{\it Superconducting state. Measured values of $\Delta V/V$ vs.~temperature for beam hits deposing in the bar an energy  $W<9$~\rm{mJ}\it. The thick dotted line is a fit to the data with a 2nd order polynomial. The thinner lines enclose a region compatible with the uncertainties in the fit parameters.}
\label{dvovfig}
\end{center}
\end{figure}

\noindent As already mentioned, the model predicts that  the amplitude maximum  of the fundamental longitudinal mode of oscillation for both states of conduction $(X_{(n)},X_{(s)})$ linearly depends on the energy deposited by the $e^-$ beam in the bar.  On the contrary, the $X_{(s)}$ values for Al~5056\cite{al2,al3} show an increasing deviation from linearity with the  increase of the deposited energy $W$, due to saturation effects discussed in Ref.~17. However, if we restrict our analysis to the data gathered at the lowest deposited energies, these non linearities can safely be neglected, and 
we can use the relation (\ref{dvov}) to compute the volume variation.  Fig.~\ref{dvovfig} shows the values of $\Delta V/V$ versus the temperature $T$ for the data with released energy:  $1.5\times 10^{-3} < W < 9\times 10^{-3}$~J.
$\Delta V/V$ data have been fitted by a 2nd order polynomial $q_0+q_1\ T+q_2\ T^2$ obtaining $q_0=(-1.737\pm 0.054)\times 10^{-8}$, $q_1=(7.334\pm 1.785)\times 10^{-9}\ \rm{K^{-1}}$, $\ q_2=(1.766\pm 0.187)\times 10^{-8}\ \rm{K^{-2}}$ with a $\chi^2=1.38$ normalized over 94 degrees of freedom. \\

The use of $q_0$ and of the relation~(\ref{Dvolume}) gives $\partial H_c/\partial P=(-3.12\pm 0.10)\times 10^{-12}\ \rm{T\ Pa^{-1}}$ at $T=0\ \rm{K}$, a value that can be compared with 
$(-2.67\pm 0.06)\times 10^{-12}\ \rm{T\ Pa^{-1}}$ obtained by Harris and Mapother in their experimental study of the critical field of pure aluminum as a function of pressure and temperature.\cite{harris}
The difference of the thermal expansion coefficients in the normal and superconducting phases is obtained by taking the derivative with respect to T of the difference of volumes:\cite{hake} $\Delta\alpha = \alpha_{(n)}-\alpha_{(s)}=\frac{1}{3}\frac{d}{dT}\frac{\Delta V}{V}$.\\
Under the hypothesis that $\alpha_{(n)}$ is still represented by the relation~(\ref{anmeas}) in the interval $0.3\ \rm{K}<T<T_c$ and by using the polynomial fitting for $\Delta V/V$, the following expression is obtained for that temperature interval:
\begin{eqnarray}
\lefteqn{\alpha_{(s)}=\alpha_{(n)}-\Delta\alpha} \nonumber \\
& =  [(-2.45\pm0.60)+(-10.68\pm1.24)T\nonumber \\
& +(0.13\pm0.01)T^3]\times 10^{-9}\ \rm{K^{-1}}
\label{asmeas}
\end{eqnarray}

\begin{figure}[htbp]
\begin{center}
\includegraphics[width=1.0\linewidth]{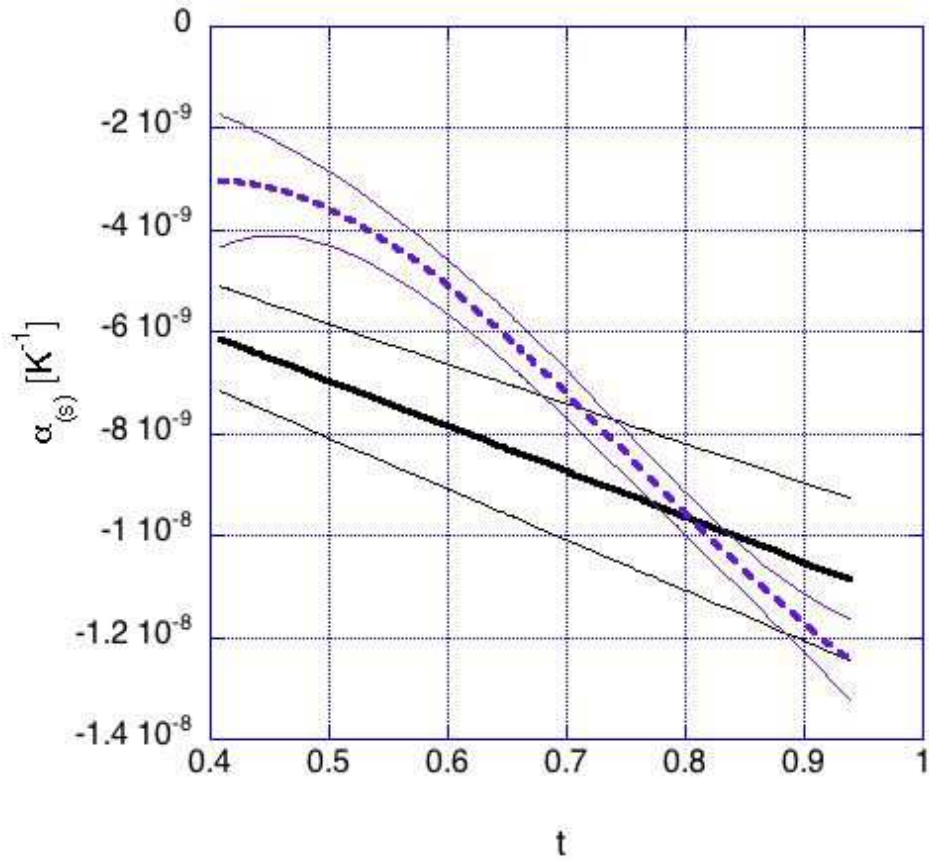}
\caption{\it Superconducting state. Values of $\alpha_{(s)}$ vs.~reduced temperature ($t=T/T_c$) for Al~5056 (solid line). For reference, we also show the result for pure Al (dotted line) derived from data of Ref.~32. The thinner lines enclose the regions compatible with the uncertainties in the $\alpha_{(s)}$ values.}
\label{alphasfig}
\end{center}
\end{figure}

\noindent This result  can be compared with the expansion coefficient of pure aluminum in the $(s)$ state, using for pure aluminum the  relation $\alpha_{(s)}=\alpha_{(s),e}+\alpha_{\ell}$, in which it is assumed that the lattice component does not depend on the conduction state and that its values are given in Ref.~33. The definition~(\ref{gru}) gives $\alpha_{(s),e}=\rho c_{V,e}{^{(s)}} \gamma_{(s),e}/(3 K_T)$, where the electronic component of $c_V$ in the $(s)$ state $(c_{V,e}{^{(s)}})$ is taken from the work of Phillips,\cite{phillips} the electronic component of the Gr\"{u}neisen parameter in the $(s)$ state $(\gamma_{(s),e})$ from Ref.~35 and $K_T^{(s)}\sim K_T^{(n)}=79.4\times 10^9\ \rm{Pa}$ at low temperatures.\cite{kamm} The comparison is shown in Fig.~\ref{alphasfig} as a function of the reduced temperature $t$.

\section{Conclusions}
In this paper we present an evaluation of the linear expansion coefficient of an Al-Mg alloy obtained by measuring both the amplitude of the fundamental mode of the longitudinal oscillation excited by electrons interacting in a suspended bar and the energy released in the bar by the electron pulse. The use of this method to determine $\alpha$ is new, although similarities exist with the direct measurement of the the Gr\"{u}neisen parameter performed by hitting thin slabs with particle beams to generate thermoelastic pulses.
No expansion data for aluminum alloys were previously available in the literature for the temperature range $0.3<T<2$~K explored by this experiment. The expansion coefficient is negative in the superconducting state and its absolute value just below $T_c$ is larger by an order of magnitude than the value above this temperature.
This feature is expected also for pure aluminum, according to an analysis\cite{marini} of the measured values of $H{_c}(P,T)$ available in Ref.~32. \\

\par
\noindent This work was partially supported by the EU FP-6 Project ILIAS (RII3-CT-2004-506222). \\

\end{document}